\documentclass[twocolumn,floatfix,preprintnumbers,nofootinbib,superscriptaddress]{revtex4}

\usepackage{ulem}
\usepackage{bm}
\usepackage{times}
\usepackage{amssymb,amsbsy,amsmath,amsfonts}
\usepackage{graphicx}
\usepackage{float}
\usepackage{color}
\usepackage{morefloats}
\usepackage{rotating}
\usepackage{srcltx}
\usepackage{slashed}
\usepackage{subfigure}
\usepackage{multirow}
\usepackage{verbatim}
\usepackage{hyperref}
\usepackage{tabularx}
\usepackage{braket}

\usepackage{CJK}

\usepackage[outdir=./]{epstopdf}

\usepackage{graphicx}
\usepackage{amsmath}
\usepackage{amsfonts}
\usepackage{amssymb}
\usepackage{color}
\usepackage{multirow}
\usepackage{mathrsfs}
\usepackage{gensymb}

\usepackage{booktabs}  
\usepackage{threeparttable}

\usepackage{subfigure}

\newcommand\be{\begin{eqnarray}}
\newcommand\ee{\end{eqnarray}}

\begin{document}

\title{The $e^+ e^- \to \Lambda^+_c \bar{\Lambda}^-_c$ cross sections and the $\Lambda_c^+$ electromagnetic form factors within the extended vector meson dominance model}

\author{Cheng Chen}~\email{chencheng@impcas.ac.cn}
\affiliation{Institute of Modern Physics, Chinese Academy of Sciences, Lanzhou 730000, China}
\affiliation{School of Nuclear Sciences and Technology, University of Chinese Academy of Sciences, Beijing 101408, China}

\author{Bing Yan}~\email{yanbing@impcas.ac.cn}
\affiliation{Institute of Modern Physics, Chinese Academy of Sciences, Lanzhou 730000, China}
\affiliation{College of Mathematics and Physics, Chengdu University of Technology, Chengdu 610059, China}

\author{Ju-Jun Xie}~\email{xiejujun@impcas.ac.cn}
\affiliation{Institute of Modern Physics, Chinese Academy of Sciences, Lanzhou 730000, China}
\affiliation{School of Nuclear Sciences and Technology, University of Chinese Academy of Sciences, Beijing 101408, China}
\affiliation{Southern Center for Nuclear-Science Theory (SCNT), Institute of Modern Physics, Chinese Academy of Sciences, Huizhou 516000, Guangdong Province, China}

\begin{abstract}

Within the extended vector meson dominance model, we investigate the $e^+ e^- \to \Lambda^+_c \bar{\Lambda}^-_c$ reaction and the electromagnetic form factors of the charmed baryon $\Lambda_c^+$. The model parameters are determined by fitting them to the cross sections  of the process $e^+e^-\rightarrow \Lambda_c^+ \bar{\Lambda}_c^-$ and the magnetic form factor $|G_M|$ of $\Lambda^+_c$. By considering four charmoniumlike states, called $\psi(4500)$, $\psi(4660)$, $\psi(4790)$, and $\psi(4900)$, we can well describe the current data on the $e^+ e^- \to \Lambda^+_c \bar{\Lambda}^-_c$ reaction from the reaction threshold up to $4.96 \ \mathrm{GeV}$. In addition to the total cross sections and $|G_M|$, the ratio $|G_E/G_M|$ and the effective form factor $|G_{\mathrm{eff}}|$ for $\Lambda^+_c$ are also calculated, and found that these calculations are consistent with the experimental data. Within the fitted model parameters, we have also estimated the charge radius of the charmed $\Lambda_c^+$ baryon. 

\end{abstract}

\maketitle
\section{Introduction}

The information about the electromagnetic structure of hadrons is reflected in the electromagnetic form factors (EMFFs)~\cite{1,2,3}. In the single photon exchange approximation, according to the four-momentum squared $q^2$ of the photon, the physical region is categorized into the spacelike region ($q^2<0$) and timelike region ($q^2>0$). Experimentally, the spacelike region corresponds to the scattering channel $e^-B \rightarrow e^-B$ ($B$ stands for a baryon), and the timelike region to the annihilation channel $e^+e^-\rightarrow B \bar{B}$ ($\bar{B}$ stands for an antibaryon). In the spacelike region, the EMFFs are associated with the charge and magnetic moment distribution of the hadrons at small momentum transfers, but they are hardly measured for the unstable hadrons, such as $\Lambda$ and $\Sigma$ with a strange quark, $\Xi$ with two strange quarks, and the $\Lambda^+_c$ with a charm quark. Therefore, the process $e^+e^-\rightarrow B \bar{B}$ is the main channel to measure the EMFFs in the timelike region for the hyperons~\cite{4, 5,6,7}.

\begin{figure}[htbp]
    \centering
    \includegraphics[scale=1.2]{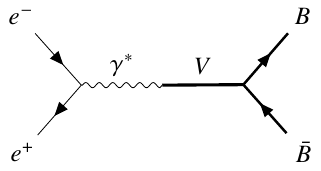}
    \caption{The Feynman diagram of $e^+e^-\rightarrow B \bar{B}$ in the VMD model.} 
    \label{fig:F}
\end{figure}

For the $e^+e^-\rightarrow B \bar{B}$ reaction, the effective form factor $|G_{\mathrm{eff}}|$ can be extracted from the Born cross section, which represents the dependence of effective coupling strength of the photon-baryon interaction vertex $\gamma B\bar{B}$, and $|G_{\mathrm{eff}}|$ is a function of the four-momentum squared $q^2$ of the virtual photon. The vector meson dominance (VMD) model is a very successful tool for studying the nucleon electromagnetic form factors, in both the spacelike and timelike regions~\cite{8,9, 10}. In the VMD model, the virtual photon coupling to baryons is through the intermediate vector mesons, as shown in Fig.~\ref{fig:F}, where $V$ stands the vector mesons that have significant couplings to the final states $B\bar{B}$. On the theoretical side, within the VMD model, the EMFFs of hyperons were investigated, such as $\Sigma$~\cite{11, 12}, $\Xi$~\cite{12}, and $\Lambda$~\cite{13,14}. 

In the charm sector, the $\Lambda_c^+$ as the lightest single-charm baryon is interesting in both theoretical and experimental studies. The Belle Collaboration firstly measured total cross sections of the process of $e^+e^-\rightarrow \Lambda_c^+ \bar{\Lambda}_c^-$, and a new charmoniumlike state, denoted as the $Y(4630)$ was discovered~\cite{15}.
After that, in 2018, the BESIII Collaboration published the precise results near the $\Lambda_c^+ \bar{\Lambda}_c^-$ threshold at four center of mass energies, and the threshold enhancement of the total cross section was found~\cite{16}. These measurements have attracted some theoretical studies. In Refs.~\cite{17, 18}, the $\Lambda^+_c \bar{\Lambda}^-_c$ final state interactions (FSI) were considered in the investigation of the $e^+e^-\rightarrow \Lambda_c^+ \bar{\Lambda}_c^-$ reaction. In addition to the $X(4660)$, the threshold enhancement was well explained by a virtual pole generated by $\Lambda^+_c \bar{\Lambda}^-_c$ attractive FSI~\cite{18}. Moreover, the EMFFs of $\Lambda_c^+$ were investigated by virtue of the VMD model in Ref.~\cite{19}, where these charmonium states $\psi(1S)$, $\psi(2S)$, $\psi(3770)$, $\psi(4040)$, $\psi(4160)$, and $\psi(4415)$ were included in the model. Note that all these above charmonium states considered in Ref.~\cite{19} are below the mass threshold of $\Lambda_c^+ \bar{\Lambda}_c^-$~\cite{20}. Hence, these nonmonotonic structures above the $\Lambda_c^+ \bar{\Lambda}_c^-$ threshold cannot be described by the model~\cite{19}.

Very recently, the BESIII Collaboration published the new measurements of the $e^+e^-\rightarrow \Lambda_c^+ \bar{\Lambda}_c^-$ reaction~\cite{21}. It is found that there is no peak around the energy of $Y(4630)$, but, a long platform behavior from the reaction threshold to $4.66\ \mathrm{GeV}$ was observed, diverging from the Belle measurements around that energy region. This new experimental result stimulates further research on the EMFFs of $\Lambda_c^+$ and the $e^+e^-\rightarrow \Lambda_c^+ \bar{\Lambda}_c^-$ reaction near threshold~\cite{22,23}. 

In the present work, based on the new measurements by the BESIII Collaboration on the $e^+e^-\rightarrow \Lambda_c^+ \bar{\Lambda}_c^-$ reaction~\cite{21}, we devoted to studying the EMFFs in the timelike region of $\Lambda_c^+$ within an extended vector meson dominance model. In the VMD model, the electromagnetic form factors of the baryons are attributed to both the couplings of the virtual photon to the vector mesons and the vector mesons to the baryons. Since the ground states $\omega$ and $\phi$ are far from the  $\Lambda^+_c \bar{\Lambda}^-_c$ mass threshold, we don't consider these two particles. On the other hand, there are a number of excited charmoniumlike states that have been experimentally discovered. For example, the $Y(4660)$ state was observed and confirmed in the $e^+e^-\rightarrow \pi^+\pi^-\psi(2S)$ reaction by the Belle, \textit{BABAR}, and BESIII Collaborations~\cite{24,25,26,27}. Moreover, a state with mass $4675.3\pm 29.5\pm3.5\ \mathrm{MeV}$ and width $218.3\pm 72.9\pm9.3\ \mathrm{MeV}$ was discovered in the process of $e^+e^-\rightarrow D^{*0}D^{*-}\pi^+$, which can be associated to the $Y(4660)$ state~\cite{28}.

In 2022, the BESIII Collaboration reported a search on the process $e^+e^- \rightarrow K^+K^-J/\psi$ at the $\rm c.m.$ energies from $4.127$ to $4.600\ \mathrm{GeV}$, and a bump structure around $\sqrt{s} = 4.5 \ \mathrm{GeV}$ was observed, which was denoted as the $Y(4500)$ state~\cite{29}. This state (called $\psi(4500)$ in this work) has also been seen in previous measurements~\cite{30, 31, 32}. Subsequently, BESIII discovered a charmoniumlike state at $\sqrt{s} = 4.7 \ \mathrm{GeV}$ with a significance over $5\sigma$ at the $K^+K^-J/\psi$ channel, denoted as $Y(4710)$ \cite{33}. The BESIII experiment has also measured the cross section of $e^+e^- \rightarrow K^0_SK^0_S J/\psi$, and found a structure analogous to $Y(4710)$~\cite{34}. In addition, a higher mass structure around $4.8$ GeV in the process $e^+e^-\rightarrow D^{*+}_sD^{*-}_s$ was found by the BESIII Collaboration~\cite{35}, denoted as $Y(4790)$. Moreover, there is a possible $\psi(4D)$ state near $4.9\ \mathrm{GeV}$, which was previously studied in Refs.~\cite{36, 37, 38}, and also, in the process $e^+e^-\rightarrow K^+K^-J/\psi$ around $5.0$ GeV appear to show a slight enhancement~\cite{31}, which might imply a state existing near the region. Here, we denoted the potential $\psi(4D)$ as $\psi(4900)$ and $Y(4790)$ as $\psi(4790)$. These charmoniumlike resonances might be related to the new findings in the $e^+e^- \to \Lambda^+_c \bar{\Lambda}^-_c$ reaction by the BESIII Collaboration~\cite{21}.

\begin{table}[htbp]
    \centering
 \caption{Masses and widths of the charmoniumlike states considered in this work.}
    \begin{tabular}{c | c | c |c}\hline\hline
        State & Mass $M_{R}$ (MeV) & Width $\Gamma_R$ (MeV) & Reference \\\hline
        $\psi(4500)$ & $4500$ & $125$ & \cite{33}  \\
        $\psi(4660)$ & $4670$ & $115$ & \cite{24} \\
        $\psi(4790)$ & $4790$ & $100$ & \cite{35} \\
        $\psi(4900)$ & $4900$ & $100$ & \cite{36, 37, 38} \\
        \hline\hline
    \end{tabular}
       \label{tab:massandwidth}
\end{table}

Based on the above discussions, we take $\psi(4500)$, $\psi(4660)$, $\psi(4790)$, and $\psi(4900)$ into account in this work. Their masses and widths are collected in Table~\ref{tab:massandwidth}.~\footnote{To minimize the model parameters, we have fixed their masses and widths.} By considering the contributions of these above excited vector states, we studied the $e^+e^- \to \Lambda^+_c \bar{\Lambda}^-_c$ reaction within the VMD model. It is found that the new measurements of the total cross sections (corresponding to the $\Lambda^+_c$ effective form factors) from threshold up $4.95 \ \mathrm{GeV}$ and the $\Lambda^+_c$ electromagnetic form factors~\cite{21} can be well described.

This article is organized as follows: in the next section, we will show the theoretical formalism for studying the $\Lambda_c^+$ electromagnetic form factors within the VMD model. Numerical results about the total cross sections of the $e^+e^-\rightarrow \Lambda_c^+ \bar{\Lambda}_c^-$ reaction, the EMFFs of $\Lambda_c^+$, and the prediction of the $\sin \Delta \Phi$ about the relative phase between electric and magnetic form factors of $\Lambda^+_c$ are shown in Sec. III. Finally, a short summary is given in Sec. IV.

\section{Theoretical formalism}

In the reaction $e^+e^-\rightarrow \Lambda^+_c \bar{\Lambda}^-_c$, within the one photon exchange approximation and based on electromagnetic current conservation and Lorentz invariance, the vertex of $\gamma \Lambda^+_c \bar{\Lambda}^-_c$ could be parameterized into two independent form factors $F_1$ and $F_2$,  which depend on the squared four-momentum transfer $q^2$ of the virtual photon. The form factors $F_1$ and $F_2$ are so-called the Pauli and Dirac form factors, respectively. Then the electric $G_E$ and magnetic form factor $G_M$ can be obtained by combining the Pauli and Dirac form factors, 
\begin{eqnarray}
    G_E(q^2) &=& F_1(q^2) + \tau F_2(q^2), \\
    G_M(q^2) &=& F_1(q^2) +  F_2(q^2),
\end{eqnarray}
where $\tau = q^2/(4M_{\Lambda^+_c}^2)$, with $M_{\Lambda^+_c}$ the mass of the charmed baryon $\Lambda^+_c$. Within the $G_E$ and $G_M$, the Born cross section of the annihilation reaction $e^+e^- \to \Lambda^+_c \bar{\Lambda}^-_c$ is given by 
\begin{eqnarray}
    \sigma  = \frac{4\pi \alpha^2 \beta C}{3s} \left( |G_M(q^2)|^2 + \frac{2M^2_{\Lambda_c^+}}{s}|G_E(q^2)|^2 \right),\label{cross}
\end{eqnarray}
where $s = q^2$ is the invariant mass squared of the $e^+ e^-$ system, and $\alpha = e^2/(4\pi)$ is the electromagnetic fine structure constant, and $\beta = \sqrt{1-4M^2_{\Lambda_c^+}/s}$ is the velocity of baryon $\Lambda_c^+$. The factor $C$ is the $S$-wave Sommerfeld–Gamow factor corresponding to the final state Coulomb interaction \cite{39}, which is: $ C(y) = \frac{y}{1-e^{-y}}$ with $y = \frac{\alpha \pi}{\beta}\frac{2M_{\Lambda^+_c}}{\sqrt{s}}$. Therefore, considering the $C$ factor, it is expected
that the cross section of the process $e^+e^-\rightarrow \Lambda^+_c \bar{\Lambda}^-_c$ is nonzero at the reaction threshold. 

According to the total cross section as in Eq.~(\ref{cross}), one can also define the effective form factor $|G_{\mathrm{eff}}|$ as
\begin{eqnarray}
    |G_{\mathrm{eff}}(q^2)| = \sqrt{\frac{2\tau|G_M(q^2)|^2+|G_E(q^2)|^2}{1+2\tau}}.
\end{eqnarray}
The effective form factor square $|G_{\mathrm{eff}}(q^2)|^2$ is a linear combination of $|G_E(q^2)|^2$ and $|G_M(q^2)|^2$, and
proportional to the total cross section of $e^+ e^- \to \Lambda^+_c \bar{\Lambda}^-_c$ reaction. Besides, the effective form factor $|G_{\mathrm{eff}}(q^2)|$ also indicates how much the experimental $e^+ e^- \to \Lambda^+_c \bar{\Lambda}^-_c$ cross section differs from a point-like charmed $\Lambda^+_c$ baryon.

In the VMD model, the virtual photon couples to $\Lambda^+_c \bar{\Lambda}^-_c$ through charmoniumlike vector mesons, thus the Dirac and
Pauli form factors are parameterized as follows:
\begin{eqnarray}
    F_1 &=& g(s)\left( f_1 +  \sum_{i=1}^4 \beta_i B_{R_i}  \right ),  \label{F1} \\
    F_2 &=& g(s)\left( f_2 B_{R_1}  + \sum_{i=2}^4 \alpha_i B_{R_i} \right), \label{F2}
\end{eqnarray}
with 
\begin{eqnarray}
B_{R_i} = \frac{M_{R_i}^2}{M_{R_i}^2 - s - i M_{R_i} \Gamma_{R_i}}, 
\end{eqnarray}
where $R_1 \equiv \psi(4500)$, $R_2 \equiv \psi(4660)$, $R_3 \equiv \psi(4790)$, and $R_4 \equiv \psi(4900)$. Their masses ($M_{R_i}$) and widths ($\Gamma_{R_i}$) are shown in Table~\ref{tab:massandwidth}. In addition, at $s = 0$ and setting the widths $\Gamma_{R_i} = 0$, with the constraints $G_E = 1$ and $G_M = \mu_{\Lambda^+_c}$, the coefficients $f_1$ and $f_2$ in Eq.~\eqref{F1} and \eqref{F2} are obtained as 
\begin{align}
    f_1 &= 1 - \beta_1 - \beta_2 - \beta_3 - \beta_4 ,\\
    f_2 &= \mu_{\Lambda_c^+} - 1 - \alpha_2 - \alpha_3 - \alpha_4 ,
\end{align}
where the $\beta_1$, $\beta_2$, $\beta_3$, $\beta_4$ and $\alpha_2$, $\alpha_3$, $\alpha_4$ are model parameters, which are determined from the fit to the current experimental data on the $e^+e^- \to \Lambda^+_c \bar{\Lambda}^-_c$ reaction. Besides, we take the magneton $\mu_{\Lambda_c^+} = 0.24\ \hat{\mu}_N$ as the theoretical prediction in Ref.~\cite{40}, which is consistent with the previous study~\cite{41}.

Furthermore, the $g(s)$ is a phenomenological intrinsic form factor that is usually chosen in the dipole form
\begin{eqnarray}
    g(s) = \frac{1}{(1-\gamma s)^2},
\end{eqnarray}
with $\gamma$ also a free parameter. In the spacelike region, taking $Q^2 = -q^2 >0$, with the dipole form for $g(Q^2)$ and the expressions of Eqs. (5) and (6) for $F_1(Q^2)$ and $F_2(Q^2)$, respectively, one can easlily find that for the large value of $Q^2$, $F_1 \sim \frac{1}{Q^4}$, and $F_2 \sim \frac{1}{Q^6}$, which are consistent with the asymptotic behavior of $F_1$ and $F_2$ calculated by the perturbative quantum chromodynamics~\cite{42,43}.

On the other hand, we will use the Flatt\'e type~\cite{44} for the $\psi(4500)$ state, since its mass is below but close to the $\Lambda_c^+\bar{\Lambda}_c^-$ threshold, and it may couple strongly to the $\Lambda_c^+\bar{\Lambda}_c^-$ channel. To do this, we take~\footnote{Only $s$-wave coupling for the $\psi(4500)$ state to the $\Lambda^+_c \bar{\Lambda}^-_c$ channel is taken into account.}
\begin{eqnarray}
    \Gamma_{\psi(4500)} = \Gamma_0 + g_{\Lambda_c}\sqrt{\frac{s}{4} - M_{\Lambda_c^+}^2},
\end{eqnarray}
with $\Gamma_0 = 125$ MeV and the second part represents the contribution from the $\Lambda_c^+\bar{\Lambda}_c^-$ channel. The $g_{\Lambda_c}$ is an unknown $s$-wave coupling constant of the $\psi(4500)$ state to the $\Lambda_c^+\bar{\Lambda}_c^-$ channel. Indeed, a $\Lambda_c^+\bar{\Lambda}_c^-$ bound state with quantum numbers $J^{PC} = 1^{--}$ is predicted in Ref.~\cite{45}, where the very near threshold data on the total cross sections of $e^+ e^- \to \Lambda_c^+\bar{\Lambda}_c^-$ reaction can be reasonably reproduced with it. While in Ref.~\cite{46}, possible $\Lambda_c^+\bar{\Lambda}_c^-$ molecular states and their productions in $p \bar{p}$ collision were investigated in a quasipotential Bethe-Salpeter equation approach.

\section{Numerical results and discussions}

Under the above formulations, we perform nine-parameter [$\gamma$, $g_{\Lambda_c}$, $\beta_i$ ($i=1,2,3,4$), and $\alpha_i$ ($i=2,3,4$)] $\chi^2$ fit to the experimental data on the total cross sections of $e^+e^- \to \Lambda^+_c \bar{\Lambda}^-_c$ reaction and the magnetic form factors $|G_M|$ of $\Lambda^+_c$ from the reaction threshold to $ 4.96\ \mathrm{GeV} $ measured by the BESIII Collaboration~\cite{16,21}. There are $28$ data points in total. The fitted parameters and errors are given in Table~\ref{tab:fittedparameters}, with a reasonable small $\chi^2/\mathrm{d.o.f} = 0.6$. The fitted result for the parameter $\gamma$ is $0.147 \pm 0.017 \ \mathrm{GeV}^{-2} $, which is smaller than the typical values for the light baryons nucleon~\cite{47,48}, $\Lambda$~\cite{14,47}, $\Sigma$~\cite{12,47}, and $\Xi$~\cite{12}.

\begin{table}[htbp]
    \setlength{\abovecaptionskip}{0cm}
    \centering
    \caption{Values of the model parameters determined in this work.}
    \begin{tabular}{cccc}
        
        \hline \hline 
         Parameter      & Value  & Parameter      & Value \\ \hline
        $g_{\Lambda_c}$ &  $1.173 \pm 0.259$    & $\beta_4$    &  $-0.141  \pm 0.097$ \\
        $\beta_1$       &  $1.883 \pm 0.484$   & $\alpha_2$    &  $1.089 \pm 0.297$ \\
        $\beta_2$       &  $-1.101 \pm 0.302$   & $\alpha_3$     &  $0.438  \pm 0.192$ \\ 
        $\beta_3$       &  $-0.439 \pm 0.194$   & $\alpha_4$     &  $0.133  \pm 0.096$ \\ \hline
    \end{tabular}
    \label{tab:fittedparameters}
\end{table}

\begin{figure}[htbp]
    \centering
    \includegraphics[scale=0.38]{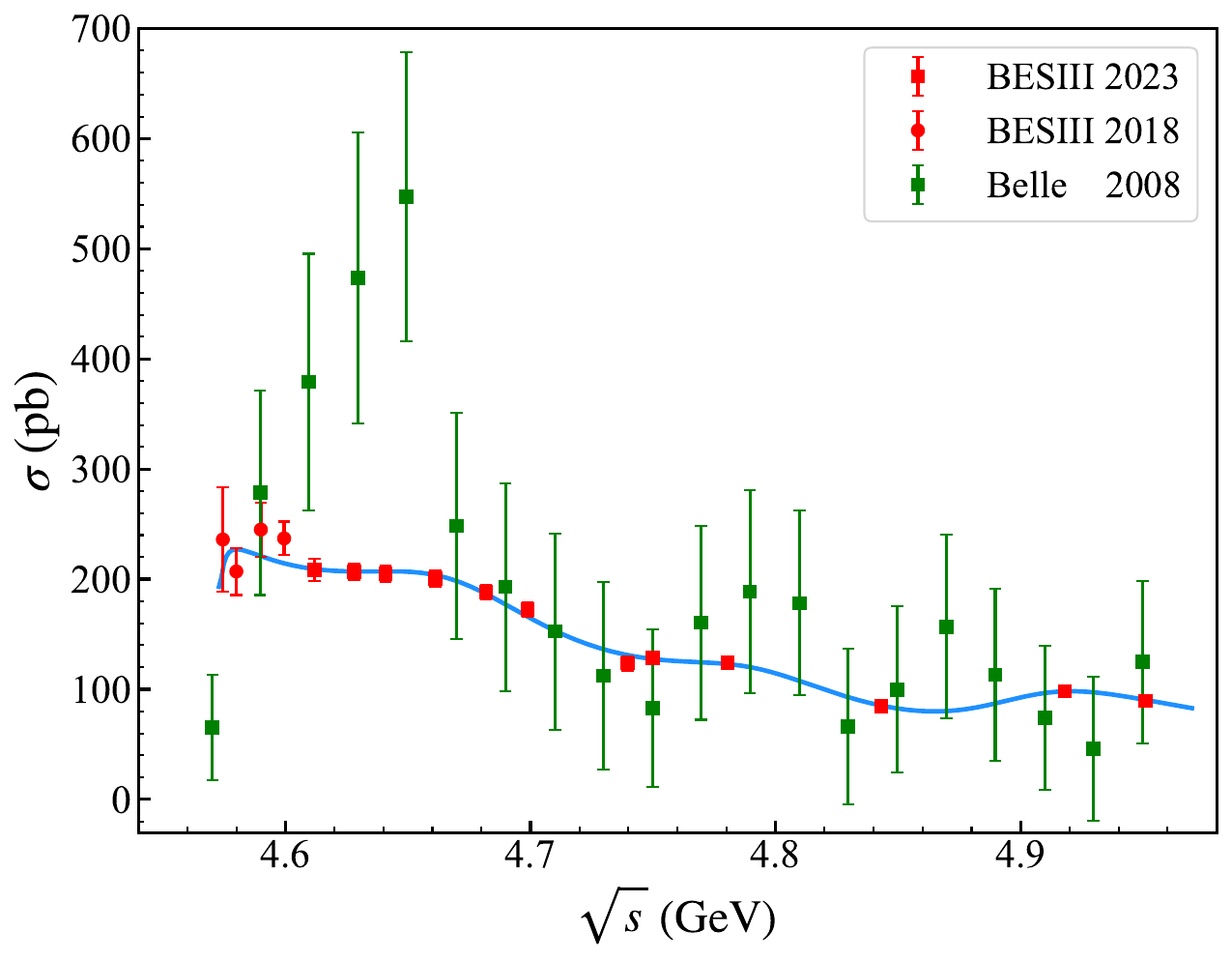}
    \caption{The obtained total cross sections compared with the experiment data. The red circle with error bars represents the data published in 2018 by the BESIII Collaboration~\cite{16}, the red rectangles in 2023 by the BESIII Collaboration~\cite{21}, and the green rectangles stand for the data taken from Belle in 2008~\cite{15}.}
    \label{fig:CS}
\end{figure}

In Fig.~\ref{fig:CS}, we show the fitted results for the total cross sections with the blue solid line compared with the experimental data.~\footnote{The data from Belle Collaboration were not taken into account in our $\chi^2$ fit.} One can see that, within the VMD model and by considering the contributions from $\psi(4500)$, $\psi(4660)$, $\psi(4790)$, and $\psi(4900)$, the total cross sections of $e^+e^- \to \Lambda^+_c \bar{\Lambda}^-_c$ reaction can be well described quite in the considered energy regions $4.57 \ \mathrm{GeV}< \sqrt{s} < 4.96\ \mathrm{GeV} $. From Fig.~\ref{fig:CS}, the cross section for the $e^+e^- \to \Lambda^+_c \bar{\Lambda}^-_c$ reaction remains almost constant from the reaction threshold to $4.67 \ \mathrm{GeV} $, exhibiting a long plateau. Here, we propose that the shape is primarily influenced by the resonances $\psi(4500)$ and $\psi(4660)$ in the VMD model. Although the structure of $\psi(4660)$ is not significant, the contribution from it is essential. Similarly, in the higher energy region, the cross sections from $4.74 \ \mathrm{GeV} $ to $4.79 \ \mathrm{GeV} $ shows a plateau with an approximate width of $50 \ \mathrm{MeV} $, attributed to the two states of $\psi(4660)$ and $\psi(4790)$. Moreover, the $\psi(4900)$ state is needed to explain the small bump structure around $4.92$ GeV.

\begin{figure}[htbp]
    \centering
    \includegraphics[scale=0.38]{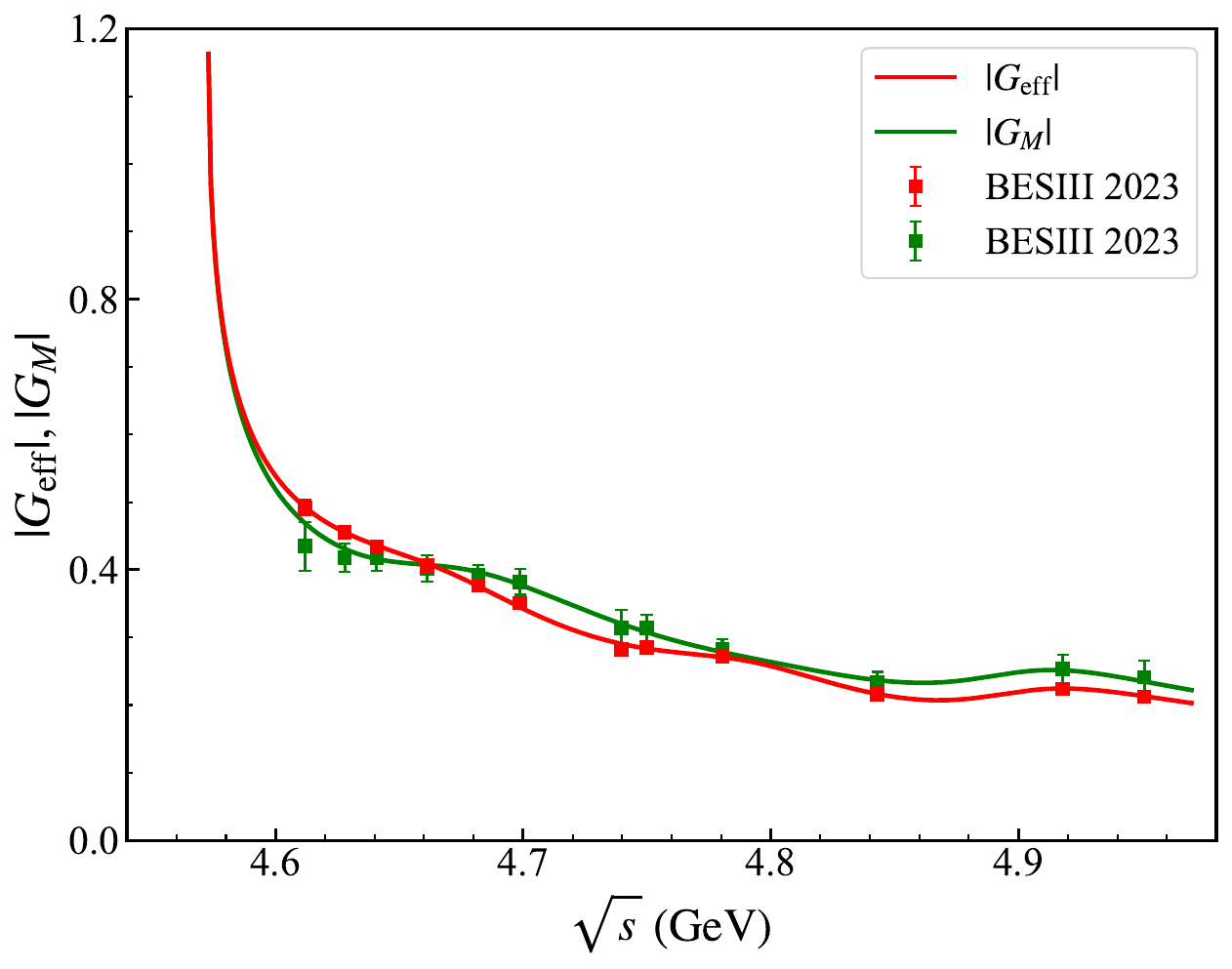}
    \caption{The obtained $\Lambda^+_c$ effective form factor $|G_{\rm eff}|$ and the fitted magnetic form factor $|G_M|$ compared with experimental data from BESIII Collaboration~\cite{21}.}
    \label{fig:GeffandGM}
\end{figure}

Since we can reproduce the total cross sections of $e^+ e^-\to \Lambda^+_c \bar{\Lambda}^-_c$ reaction, it is expected that the $\Lambda^+_c$ effective form factor can also be described well, as shown in Fig.~\ref{fig:GeffandGM} by the red line, compared with the experimental data~\cite{21}. In Fig.~\ref{fig:GeffandGM}, we also show the fitted numerical results for the $\Lambda^+_c$ magnetic form factor $|G_M|$ (the green line) compared with the experimental measurements. Again, one can see that the fitted results are in agreement with the experimental measurements of $|G_M|$ by the BESIII Collaboration~\cite{21}. This indicates that the nonmonotonic line shapes of the $e^+e^-\to \Lambda^+_c \bar{\Lambda}^-_c$ cross sections and the $\Lambda^+_c$ effective form factors can be explained within the VMD model, where the contributions of the charmoniumlike states are taken into account.

The effective form factor $|G_{\mathrm{eff}}|$ for $\Lambda^+_c$, as stated in Ref.~\cite{21}, does not exhibit oscillatory behavior, in contrast to the findings in the nucleon case~\cite{47,49,50,51,52,53,54}. This may be attributed to the limited energy range considered for the $e^+ e^- \to \Lambda^+_c \bar{\Lambda}^-_c$ reaction. The range from the reaction threshold to 4.95 GeV is only about 350 MeV. More comprehensive experimental data covering wider energy regions are required to further confirm the so-called oscillatory behavior of the effective form factor $|G_{\mathrm{eff}}|$ for $\Lambda^+_c$. Nonetheless, nonmonotonic structures in the line shape of $\Lambda^+_c$ effective form factors are indeed present, and these can be naturally reproduced within the vector meson dominance model by considering the contributions from excited vector states.

As discussed above, the Coulomb enhancement factor leads to a nonzero cross section for the process $e^+e^- \to \Lambda^+_c \bar{\Lambda}^-_c$ at the reaction threshold $s = 4M^2_{\Lambda^+_c}$. Utilizing the fitted parameters shown in Table~\ref{tab:fittedparameters}, the $e^+e^- \to \Lambda^+_c \bar{\Lambda}^-_c$ cross section at the threshold can be obtained  
\begin{align}
    \sigma_{\rm thre.} = \frac{\pi^2 \alpha^3}{2 M_{\Lambda^+_c}^2}|G_{\mathrm{eff}}(4 M^2_{\Lambda^+_c})|^2 = 193.1 \ \mathrm{pb},
\end{align}
and the corresponding $\Lambda^+_c$ effective form factor is $|G_{\mathrm{eff}}(4 M^2_{\Lambda_c^+})| = 1.16$, which is larger than the value of $|G_{\mathrm{eff}}(4 M^2_{p})| = 1.00\pm 0.05$ for the case of proton~\cite{2}. This may indicate that the Coulomb interaction dominates the $e^+ e^- \to p \bar{p}$ reaction near the threshold~\cite{56,55}. However, for the case of the charmed $\Lambda_c^+$ baryon, the contribution from the strong interaction is prominent.

\begin{figure}[htbp]
    \centering
    \includegraphics[scale=0.38]{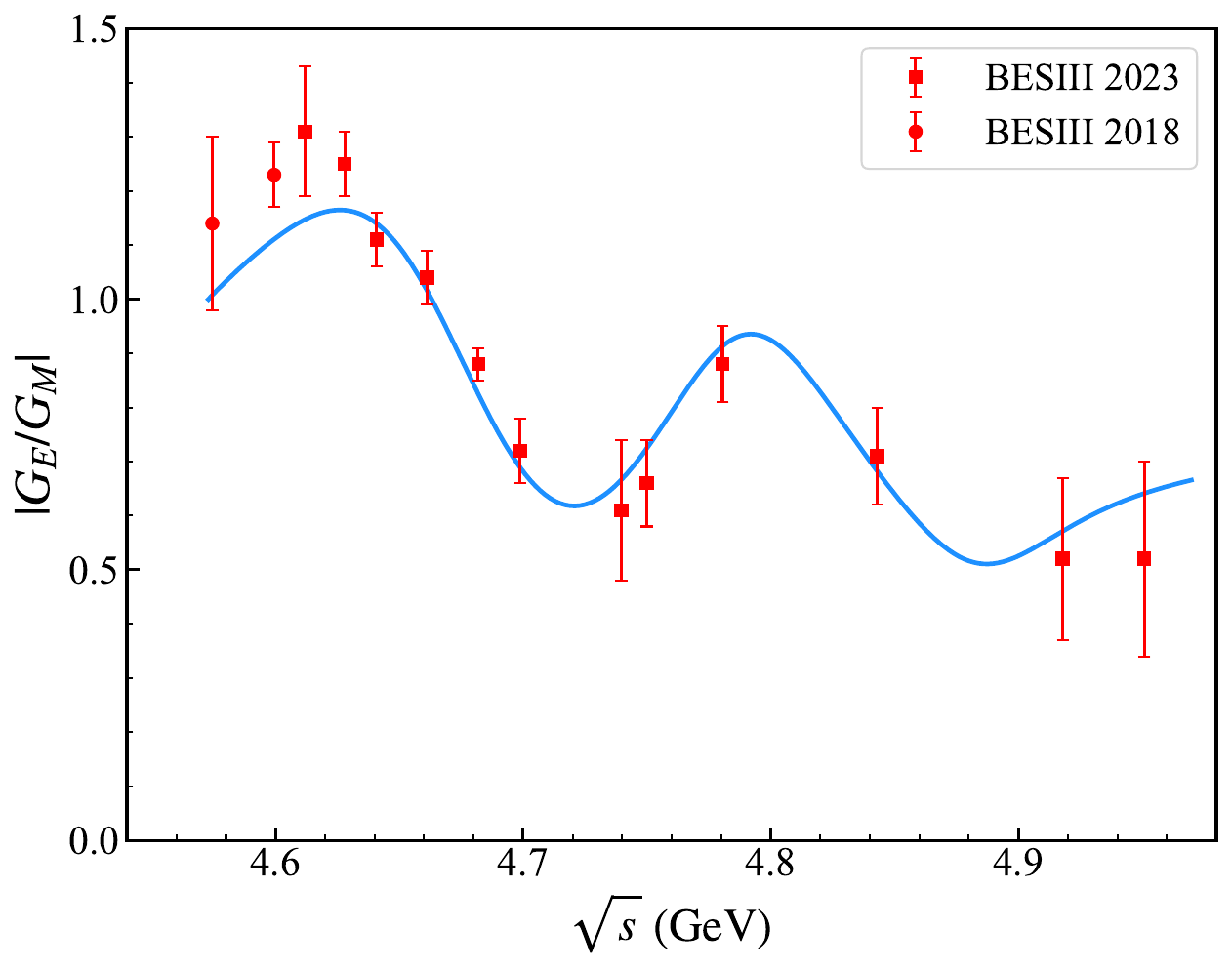}
    \caption{ The ratio $|G_E/G_M|$ compared with experimental data from BESIII \cite{16, 21}.}
    \label{fig:Ratio}
\end{figure}

Within the fitted parameters in Table~\ref{tab:fittedparameters}, we have calculated the ratio $|G_E/G_M|$, and it is shown in Fig.~\ref{fig:Ratio}. One can clearly see that the obtained $|G_E/G_M|$ are in good agreement with experimental data. The non-monotonic structures (or the so-called oscillating behavior as in Ref.~\cite{21}) shown in the line shape of the ratio $|G_E/G_M|$ can be naturally explained by including the contributions from the charmonium states $\psi(4500)$, $\psi(4660)$, $\psi(4790)$, and $\psi(4900)$. Note that the $\psi(4900)$ state is crucial to reproduce the experimental measurements on the $e^+e^-\rightarrow \Lambda_c^+ \bar{\Lambda}_c^-$ reaction, even there are a few data points in the energy regions from $4.85\ \mathrm{GeV}$ to $4.96\ \mathrm{GeV}$. It is expected that more experimental measurements around $4.9$ GeV can be used to further study the possible $\psi(4900)$ state~\cite{57}.

\begin{figure}[htbp]    
    \centering
    \includegraphics[scale=0.38]{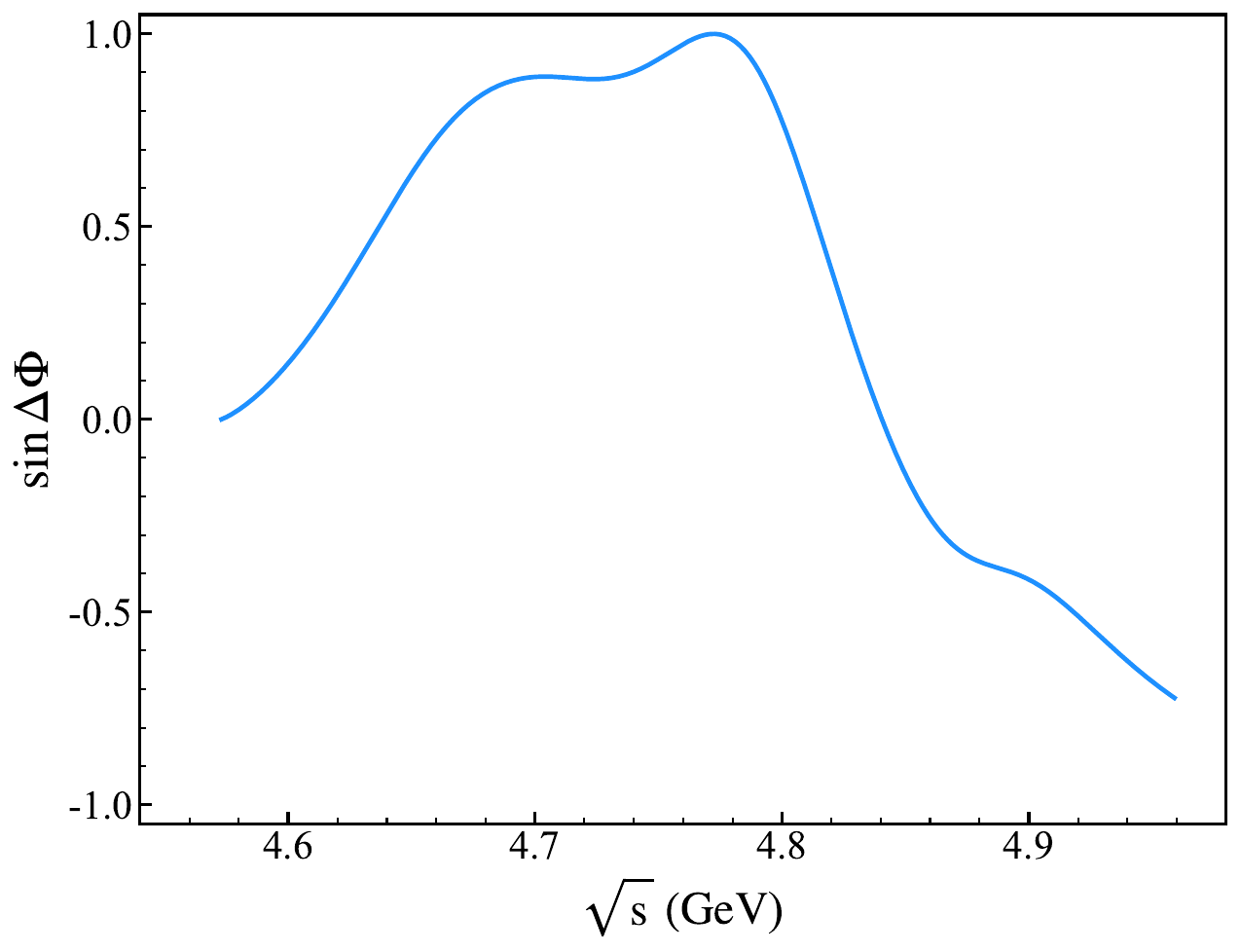}
    \caption{The predictions for $\sin \Delta \Phi$ as a function of $\sqrt{s}$.}
    \label{fig:Sin}   
\end{figure}

In the timelike region, the electric $G_E$ and magnetic form factor $G_M$ are complex. The relative phase $\Delta \Phi$ between them is associated with the spin polarization of the $\Lambda^+_c$ in the process of $e^+e^-\rightarrow \Lambda_c^+ \bar{\Lambda}_c^-$~\cite{58, 59, 60}. Here, we can write
\begin{align}
    G_E/G_M  = e^{i \Delta \Phi} |G_E/G_M|.
\end{align}
Using the parameters in Table~\ref{tab:fittedparameters}, we can obtain the functional dependence of $\sin \Delta \Phi$ on $\sqrt{s}$, which is shown in Fig.~\ref{fig:Sin}. At the threshold, $\sin \Delta \Phi$ is 0 due to the equality of $G_E$ and $G_M$, subsequently increasing to higher values. From  $4.68$ to $4.78 \ \mathrm{GeV} $, the values of $\sin \Delta \Phi $ are larger than $0.8$. This suggests that 
the larger polarizations of $\Lambda^+_c$ are more likely to be experimentally observed in this energy region. In fact, given the total cross sections of the $e^+ e^- \to \Lambda^+_c \bar{\Lambda}^-_c$ reaction as shown in Fig.~\ref{fig:CS}, we propose that it would be more efficient to conduct polarization measurements at the center-of-mass energies from $4.68$ to $4.70 \ \mathrm{GeV} $. 

It is important that the $\sin \Delta \Phi$ reaches 0 again at around $\sqrt{s} = 4.84 \ \mathrm{GeV}$ and then decreases steadily to large negative values, which might also imply a considerable polarization of $\Lambda^+_c$. However, it is worthy nentioning that only the data in the range $4.57 < \sqrt{s} < 4.96\ \mathrm{GeV} $ were included in our fitting, and thus we do not have the information on the results for the $\sin \Delta \Phi$ at the higher energy region.

\begin{figure}[htbp]  
    \centering
    \includegraphics[scale=0.38]{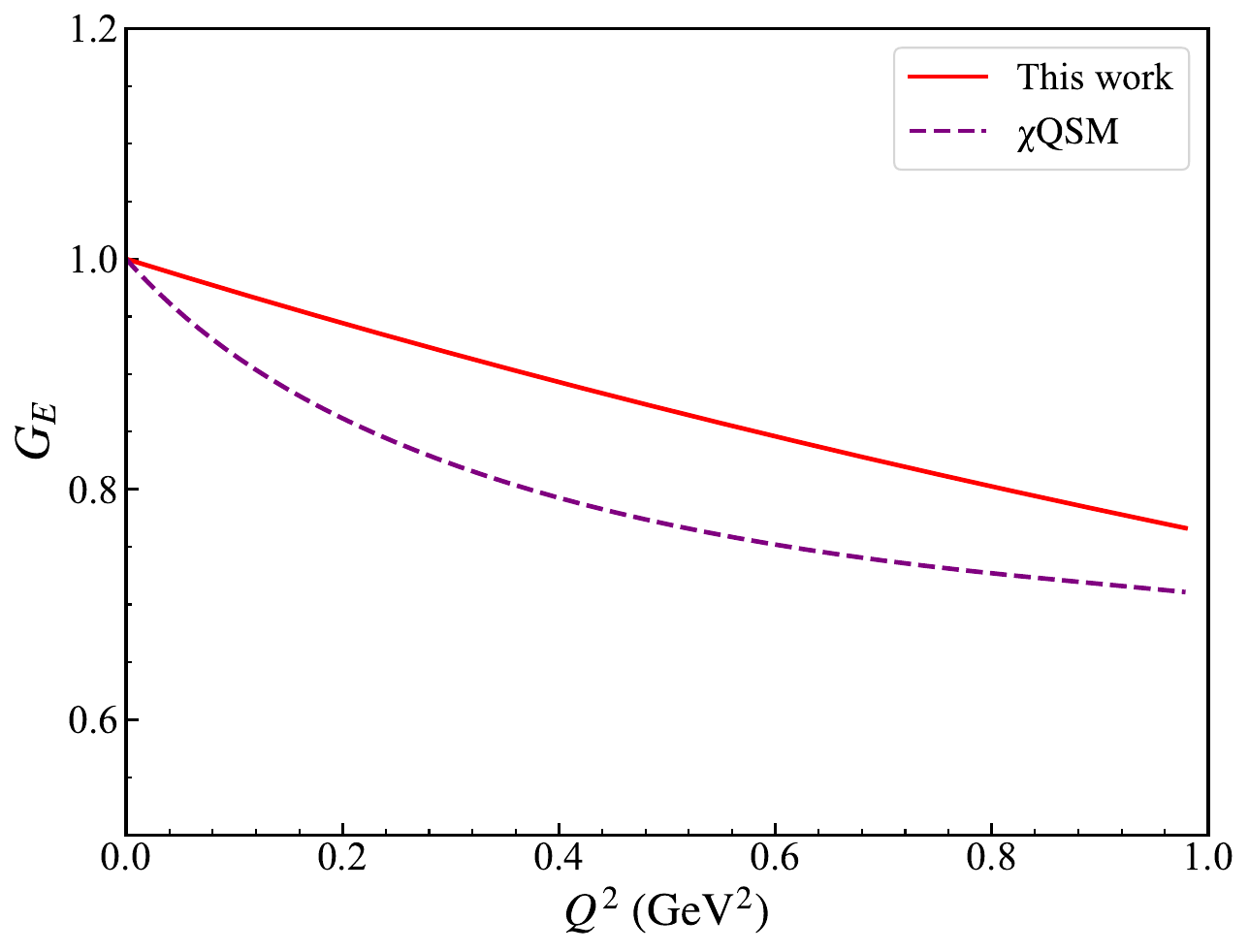}
    \caption{ The red line is the prediction of electric form factor $G_E$ in the spacelike region. The purple dashed line is the results in Ref.~\cite{61}.}
    \label{fig:GEspce}
\end{figure}

Next, we turn to the electromagnetic form factors in the spacelike region.~\footnote{Note that turning to the spacelike region, these low-lying excited vector states should also contribute to the $\Lambda^+_c$ elactromagnetic form factors in the $t$-channel exchanges. Such kind of mechanism is checked in the present work based on the insufficient experimental data of $e^+ e^- \to \Lambda^+_c \bar{\Lambda}^-_c$ reaction in the timelike region, and it is found that the contributions from the low-lying excited vector states could be very small and can be neglected.} The spacelike electromagnetic form factors can be easily obtained by converting $q^2$ to $-Q^2$ and setting the widths $\Gamma_{R_i} \ (i = 1,2,3,4)$ in Eqs.~(\ref{F1}) and (\ref{F2}) to zero. The obtained $\Lambda^+_c$ electric form factor $G_E$ in the spacelike region is shown in Fig.~\ref{fig:GEspce}, where theoretical calculations in Ref.~\cite{61} are also shown for comparison. Our results are somewhat different quantitatively from the calculations in Ref.\cite{61} with the self-consistent SU(3) chiral quark-soliton model. It is expected that these theoretical calculations can be tested by future experiments on the EMFFs of the charmed $\Lambda^+_c$ baryon and thus will provide new insight into the production of $\Lambda^+_c$ in the $e^+ e^- \to \Lambda^+_c \bar{\Lambda}^-_c$ reaction.

Then one can also calculate the $\Lambda_c^+$ mean squared charge radius, which is given by
\begin{align}
    {\left\langle r_{\mathrm{E}}^2  \right\rangle}_{\Lambda_c^+} = \frac{-6}{G_E(0)}{\frac{\mathrm{d}G_E(Q^2)}{\mathrm{d}Q^2}}\bigg|_{Q^2=0} . \label{radius}
\end{align} 
This dictates the charge radius being determined by the slope of $G_E$ at $Q^2 = 0$, and $G_E(0) = 1$ corresponds to the charge value of $\Lambda_c^+$. The radius is determined to be ${\left\langle r_{E}^2  \right\rangle}_{\Lambda_c^+} = 0.07 \ \mathrm{fm}^2$ using the Eq.~(\ref{radius}) and the parameters listed in Table~\ref{tab:fittedparameters}. The value obtained here is much smaller than the value, ${\left\langle r_{E}^2  \right\rangle}_{\Lambda_c^+} = 0.24 \ \mathrm{fm}^2$, obtained in Ref.~\cite{61}. This difference can be easily seen from the two curves shown in Fig.~\ref{fig:GEspce}.

\section{Sumarry}

In this work, we investigate the electromagnetic form factors of charmed $\Lambda_c^+$ baryon and the total cross sections of the $e^+e^- \to \Lambda^+_c \bar{\Lambda}^-_c$ reaction within the extended vector meson dominance model. The model parameters are determined by fitting them to the experimental data on the $e^+e^- \to \Lambda^+_c \bar{\Lambda}^-_c$ total cross sections and the $\Lambda^+_c$ magnetic form factor $|G_M|$, as measured by the BESIII Collaboration~\cite{16, 21}. By including the charmoniumlike states $\psi(4500)$, $\psi(4660)$, $\psi(4790)$, and the possible $\psi(4900)$ state, the current experimental data can be well reproduced. Notably, the threshold enhancement of the total cross sections of the $e^+e^- \to \Lambda^+_c \bar{\Lambda}^-_c$ reaction can be explained by the model, primarily attributed to the $\psi(4500)$ and $\psi(4660)$ states. Additionally, the possible $\psi(4900)$ state is crucial to reproduce the small bump structure around the center-of-mass energy $\sqrt{s} = 4.92$ GeV.

Using the fitted model parameters, we calculated the ratio $|G_E/G_M|$ and effective form factor $|G_{\mathrm{eff}}|$, and found that these calculations are consistent with the experimental data. Moreover, the relative phase $\Delta \Phi$ about $G_E$ and $G_M$ has been calculated in our work. We directly provided the numerical results of $\sin \Delta \Phi$, and found that it is more efficient to measure the spin polarization of $\Lambda^+_c$ in the $e^+ e^- \to \Lambda^+_c \bar{\Lambda}^-_c$ reaction around the center-of-mass energy $\sqrt{s} = 4.69 \ \mathrm{GeV}$.

Finally, we also studied the $\Lambda^+_c$ electromagnetic form factors in the spacelike region using the parameters obtained from the fit in the timelike region. According to the results of $G_E$, the mean squared charge radius is determined to be $0.07 \ \mathrm{fm}^2 $. The theoretical calculations here can be tested by future experiments and further experimental measurements will provide new insight into the production of $\Lambda^+_c$ in the $e^+ e^- \to \Lambda^+_c \bar{\Lambda}^-_c$ reaction.

In conjunction with the studies in Ref.~\cite{14} for the $\Lambda$, in Ref.~\cite{12} for the $\Sigma$ and $\Xi$, and in Ref.~\cite{48} for the nucleon, we conclude that the non-monotonic structures observed in the line shape of the $e^+ e^- \to B\bar{B}$ total cross sections can be naturally explained within the vector meson dominance model. Additionally, the $e^+ e^- \to B\bar{B}$ reactions can be used to study the excited vector states, especially for those exotic states that couple strongly to the $B\bar{B}$ channels.

\section*{Acknowledgements}

This work is partly supported by the National Key R\&D Program of China under Grant No. 2023YFA1606703, and by the National Natural Science Foundation of China under Grant No. 12075288. It is also supported by the Youth Innovation Promotion Association CAS.


\end{document}